%% file: MBNS_proceedings.tex
\documentclass[a4paper]{jpconf}
\usepackage{graphicx}
\usepackage{graphics}
\usepackage{amssymb}
\usepackage{amsmath}
\usepackage{citesort}
\usepackage{comment} 

\input kigou.tex
\begin{document}

\title{Quasi-equilibrium models of magnetized compact objects}

\author{
  Charalampos Markakis${}^1$,
  K\=oji Ury\=u${}^{2}$ and 
  Eric Gourgoulhon${}^3$ 
}
\address{
${}^1$Department of Physics, University of Wisconsin-Milwaukee, P.O. Box 413,  
Milwaukee, WI 53201, USA\\
${}^2$Department of Physics, University of the Ryukyus, Senbaru, 
Nishihara, Okinawa 903-0213, Japan
\\
${}^3$Laboratoire Univers et Th\'eories, UMR 8102 du CNRS,
Observatoire de Paris, Universit\'e Paris Diderot, F-92190 Meudon, France
\\
}

\ead{markakis@uwm.edu${}^1$, uryu@sci.u-ryukyu.ac.jp${}^{2}$, eric.gourgoulhon@obspm.fr${}^3$ }

\begin{abstract}
We report work towards a relativistic formulation for modeling strongly
magnetized neutron stars, rotating or in a close circular orbit around another neutron star
or black hole, under the approximations of helical symmetry and ideal MHD.
The quasi-stationary evolution is governed by the first law of thermodynamics for
helically symmetric systems,  which is generalized to include magnetic fields. The
formulation involves an iterative scheme for solving the Einstein-Maxwell and
relativistic MHD-Euler equations numerically. The resulting configurations for binary
systems could be used as self-consistent initial data for studying their inspiral and
merger.
\end{abstract}

\section{Introduction}
A uniformly rotating neutron star is significantly deformed
when the ratio of  kinetic energy 
$T$ to  gravitational energy $W$ becomes 
$T/|W|\sim 0.1$.  
When the magnetic field energy, 
%
\beq
{\cal M} \,:=\,  \frac1{8\pi} \int B^2 d^3x,
\eeq
becomes such that ${\cal M}/|W| \sim 0.01$, the contribution 
of the magnetic field to the structure of the neutron srar 
may not be neglected. This is estimated to occur for 
\beq
B\sim 4.4\times 10^{16}
\left(\frac{M \,[M_\odot]}{1.4 \,[M_\odot]}\right)
\left(\frac{10 \,\mbox{[km]}}{R \,\mbox{[km]}}\right)^2
\mbox{[G]}.
\eeq
Recent observations of anomalous X-ray pulsars, or soft $\gamma$-ray 
repeaters suggest that the neutron stars in these systems 
may be associated with strong magnetic fields around $10^{14}$ -- $10^{15}$ G 
at the surface (see, e.g. \cite{2006csxs.book..547W}).  
We may expect that the interior magnetic field of 
such a strongly magnetized neutron star, a magnetar,  
is a few orders of magnitudes stronger.

Such strong magnetic fields have not been found in binary neutron 
star systems, and may not survive until merger.  
Hypothetically, however, strongly magnetized neutron stars 
or black holes may form binary neutron star or 
black hole - neutron star systems.  For instance, 
the poloidal field seen as the surface magnetic field 
decays but the toroidal field stays strong enough 
to affect the structure of the compact objects.

In several numerical relativity simulations of magnetized binary 
neutron stars, it has been shown that, 
after  binary inspiral and merger, the magnetic fields are amplified around 10 
times by  magnetic winding and the magnetorotational instability 
in post-merger, pre-collapse objects \cite{BNSmagnetized}.  
Such an object is likely to form a black hole and a magnetized 
toroid system, which becomes the source of a short $\gamma$-ray burst. 
It is desirable to prepare realistic initial data sets for such 
merger simulations calculated by solving the Einstein-Maxwell equations 
and a first integral of the MHD-Euler equation, assuming 
(quasi-)equilibrium.  
In this article, we model such magnetized binary compact objects 
in close circular orbits, assuming that the spacetime and magnetic fields 
satisfy  helical symmetry and that the stars are in equilibrium 
\cite{BD92D94,schild,FU06GU07}. 

\section{Thermodynamic laws for helically symmetric
Einstein-Maxwell spacetimes
with charged and magnetized perfect fluids}

\subsection{Zeroth law}

We consider a globally hyperbolic spacetime $({\cal M}, \gabd)$.  
The field $k^\alpha$ is transverse to each 
Cauchy surface, not necessarily timelike everywhere, 
and generates a one-parameter family of 
diffeomorphisms $\chi_t$.  The action of $\chi_t$ 
on a spacelike sphere $\cal S$ on a Cauchy surface 
generates a timelike surface, 
${\cal T}({\cal S}) = \cup_t \chi_t({\cal S})$, called 
the \emph{history of $\cal S$}.  
Then, $k^\alpha$ is a \emph{helical vector} field if 
there is a smallest $T>0$ for which $P$ and $\chi_T(P)$ are 
timelike separated for every point $P$ outside of 
the history ${\cal T}({\cal S})$.  
A vector $k^\alpha$ written as 
\beq
k^\alpha = t^\alpha + \Omega \phi^\alpha
\label{eq:hkv}
\eeq
is the helical vector, where $t^\alpha$ is a timelike vector 
and $\phi^\alpha$ a spacelike vector that has circular orbits 
with a parameter length $2\pi$ (see, FUS).  In Fig.~1, 
a schematic figure of the history ${\cal T}({\cal S})$ generated 
by the diffeo $\chi_t$ is presented.

\begin{figure}
\begin{tabular}{cc}
\begin{minipage}{.55\hsize}
\begin{center}
\includegraphics[height=40mm]{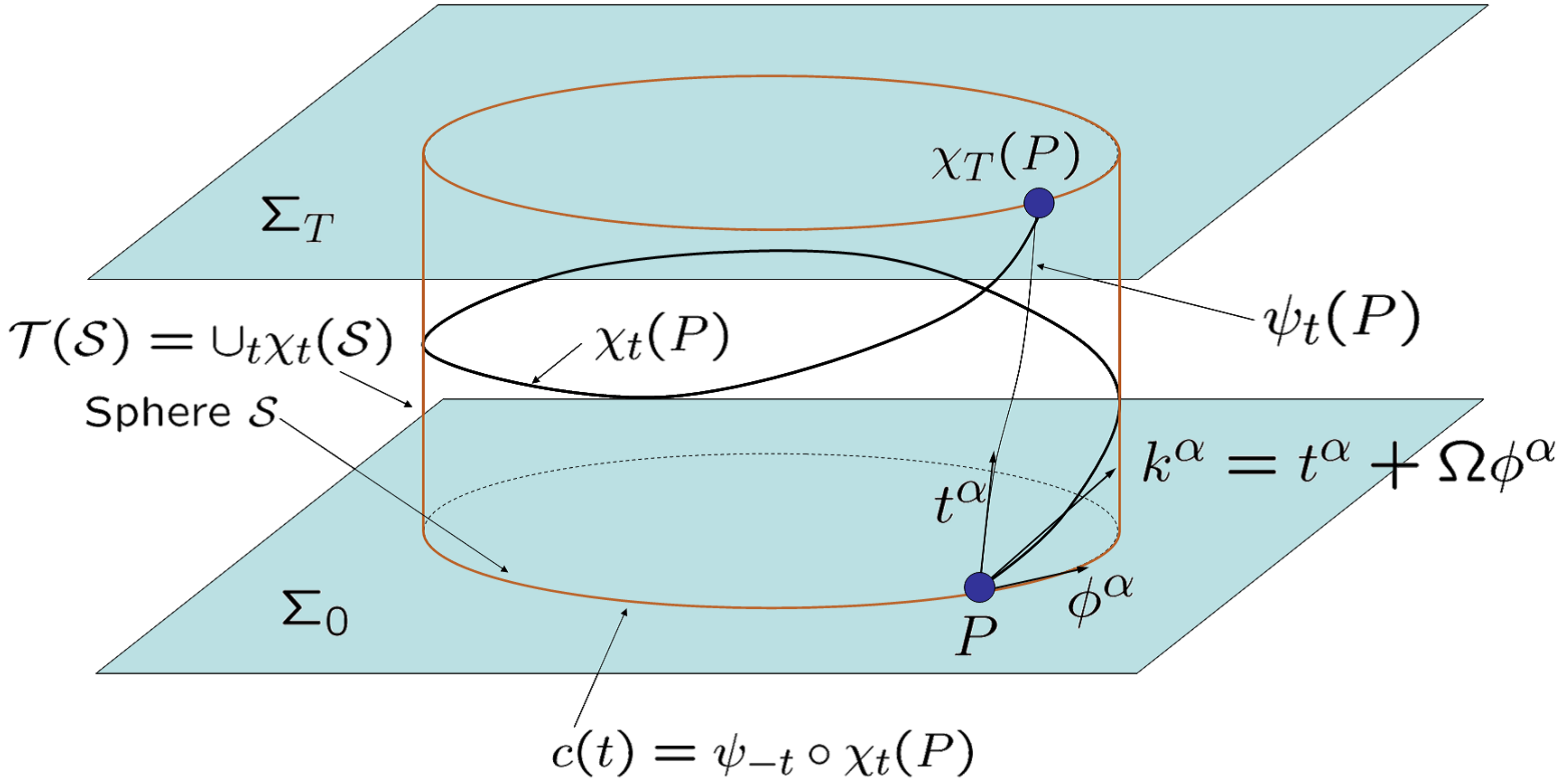}
\end{center}
\end{minipage} 
&
\begin{minipage}{.45\hsize}
\begin{center}
\includegraphics[height=45mm]{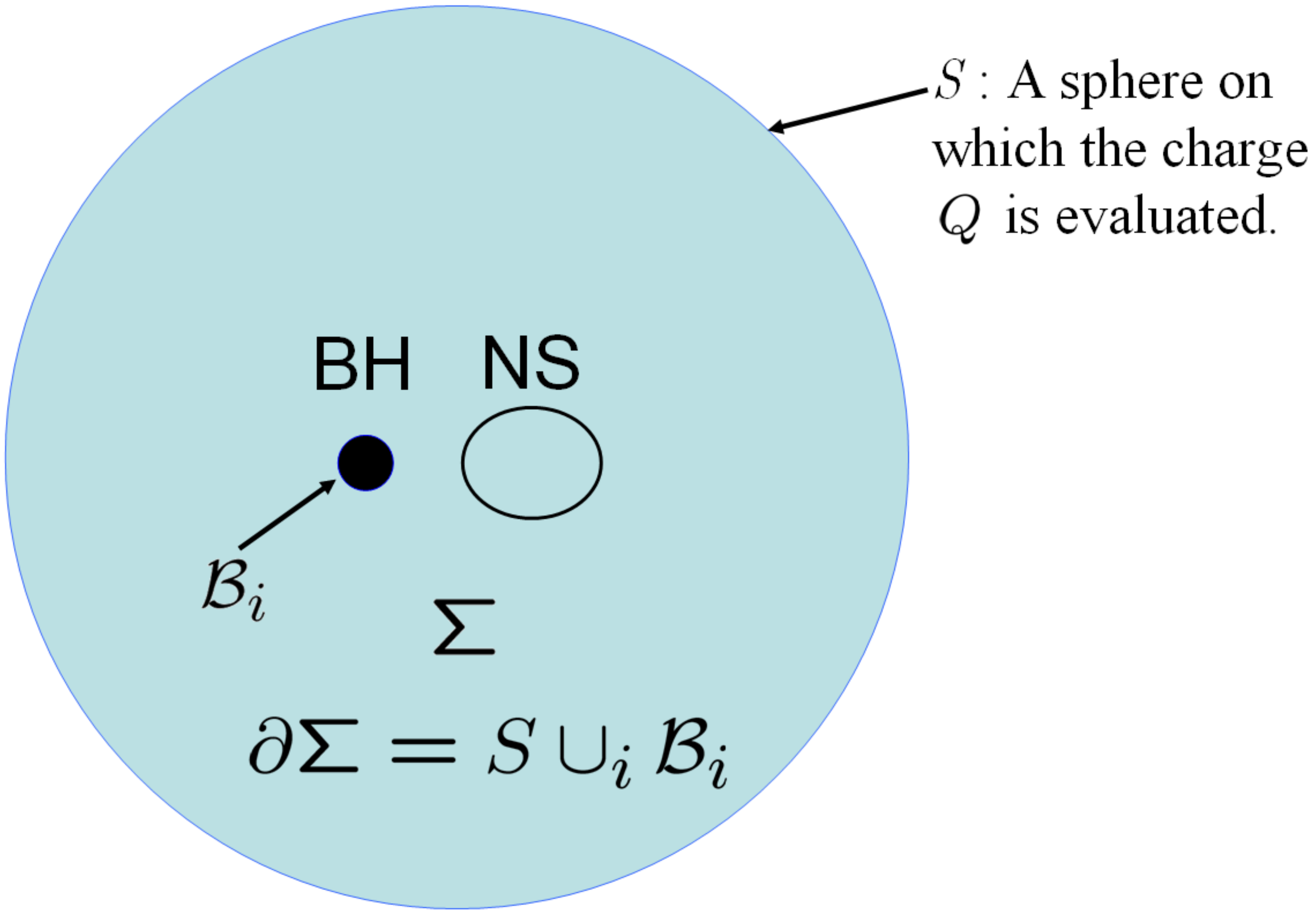}
\end{center}
\end{minipage} 
\end{tabular} 
\caption{Left panel: the history of a sphere $\cal S$ 
generated by the helical vector $k^\alpha$.
Right panel: a domain of integration $\Sigma$ and its boundaries.}
\end{figure}

Each Cauchy surface of a helically symmetric 
spacetime does not admit  flat asymptotics.  
Therefore, the future (past) horizon $\Hori$ is defined as 
the boundary of the future(past) domain of outer communication 
${\cal D}^\pm$ of a history ${\cal T}({\cal S})$ of each 
spacelike sphere $\cal S$.  
If the history 
${\cal T}({\cal S})$ of a sphere $\cal S$ is in 
${\cal D}^\pm$, the future (past) horizon agrees 
with the chronological past (future) of the history 
${\cal T}$, $\Hori = \pa I^\mp({\cal T})$.

The existence of a global helical symmetry assures that the horizon 
is a Killing horizon.  
With the null energy condition $R_\albe l^\alpha l^\beta \ge 0$
for any null vector $l^\alpha$, the surface gravity $\kappa$, 
defined on each connected component of the horizon $\Hori$ by 
\beq
k^\beta \na_\beta k^\alpha \,=\, \kappa k^\alpha ,
\eeq
is constant.  The proof is given in FUS, in which the
conditions of theorems by Friedrich, R\'acz, and Wald \cite{FRW} 
are modified to make them suitable for helically symmetric spacetimes.
Also the electric potential $\PhiE$
in the rotating frame, defined by 
\beq
E_\alpha \,=\, F_\albe k^\beta \,=\, - \na_\alpha \PhiE, \qquad 
\PhiE \,=\, A_\alpha k^\alpha + {\rm const},
\label{eq:PhiE}
\eeq
is constant on $\Hori$, assuming the symmetry $\Lie_k A_\alpha =0 $. 
Since $E_\alpha$ is null on $\Hori$ and 
$E_\alpha k^\alpha = 0$, $E_\alpha$ is parallel to the null generator 
on  $\Hori$.  Hence for any vector $\eta^\alpha$ tangent to $\Hori$, 
$\eta^\alpha E_\alpha = -\eta^\alpha \na_\alpha \PhiE = 0$.  
Therefore, as for the stationary and axisymmetric spacetimes 
shown by Carter \cite{Carter73,Carter79}, the surface gravity $\kappa$ and the 
electric potential $\PhiE$ in the rotating frame  are constant 
on the horizon $\Hori$.

\subsection{First law}

Consider a family of spacetimes, 
\beq
{\cal Q}(\lambda) := [g_{\alpha\beta}(\lambda), u^\alpha(\lambda),  
\rho(\lambda), s(\lambda), A_\alpha(\lambda), j^\alpha(\lambda)],   
\label{eq:Qvars}
\eeq
whose Lagrangian density is written 
\beq
\Lag \,=\,
\left(\frac1{16\pi} R \,-\, \epsilon 
\,-\,\frac1{16\pi} \Fabd \Fabu \,+\, A_\alpha j^\alpha
\right)\sqrt{-g}, 
\label{eq:Lag}
\eeq
where 
$u^\alpha$, $\rho$, $s$, $\epsilon$, and $j^\alpha$ are, respectively, 
the 4-velocity, baryon rest mass density, entropy per baryon mass, 
energy density, and  electric 4-current.  
A generalized first law for the spacetime ${\cal Q}(\lambda)$ associated 
with the helical symmetry is derived as 
a variation formula of the Noether charge associated with 
the helical Killing vector defined by \cite{wald,iyerwald} 
\beq
Q(\lambda) = \oint_S Q^{\alpha\beta}\dSab,
\label{eq:Q}
\eeq
\beq
\hB^\alpha (\lambda) 
\,=\, \frac1{16\pi}(g^{\alpha\gamma}g^{\beta\delta} 
                 -g^{\alpha\beta}g^{\gamma\delta})|_{\lambda=0}  
                 \zna_\beta g_{\gamma\delta}(\lambda) 
\,+\,\frac1{4\pi}F^{\beta\alpha}|_{\lambda=0}
\Big[A_\beta(\lambda)-\frac12 A_\beta(0)\Big]
\,+\,\rm O(\lambda^2),  
\eeq
With this choice of $\hB^\alpha$, the charge $Q(\lambda)$ 
becomes finite and independent of the sphere $S$ used to evaluate 
the charge $Q(\lambda)$ as long as $S$ encloses all black holes and neutron stars.

To see this, we first check that $Q(0)$ is independent of $S$.  
The surface integral of (\ref{eq:Q}) is rewritten in terms of 
integrals over a spacelike hypersurface $\Sigma$ transverse to 
$k^\alpha$ and over the black hole boundary ${\cal B}_i$, 
where ${\cal B}_i$ is the $i$-th connected component of $\Sigma \cap \Horiplus$.  
In Fig.1, the region of integration is shown.  
Since, the boundary of $\Sigma$ is the union of the sphere $S$ and 
${\cal B}_i$, $\pa\Sigma = S \cup_i {\cal B}_i$, 
a difference $Q\,-\,\sum_i Q_i$ is calculated at $\lambda = 0$: 
\beqn
&&
Q\,-\,\sum_i Q_i 
\,=\, - \frac{1}{8\pi} \int_\Sigma (\Gab k^\beta - 8\pi \TabF) dS_\alpha 
        - \frac{1}{16\pi} \int_\Sigma R k^\alpha dS_\alpha. 
\nonumber\\
&&\qquad
\,+\, \int_{\Sigma} 
\left(\frac1{8\pi}\na_\gamma F^{\beta\gamma} A_\beta k^\alpha
\,-\,\frac1{4\pi} k^\gamma A_\gamma \na_\beta \Fabu  \right) dS_\alpha
-\,\sum_i
\frac1{4\pi} \int_{{\cal B}i}
k^\gamma A_\gamma F^{\alpha\beta} dS_\albe .  
\label{eq:calcQ}
\eeqn
where $Q_i(\lambda)$ is defined on ${\cal B}_i$ by 
\beq
Q_i(\lambda) := \oint_{{\cal B}_i} Q^{\alpha\beta}\dSab,
\label{eq:Qi}
\eeq
$\TabF$, the stress-energy tensor of 
the electromagnetic field, is defined by,   
\beq
\TabFu\,=\,
\frac1{4\pi}\left(F^{\alpha\gamma}F^\beta{}_\gamma
-\frac14\gabu\Fcdd\Fcdu\right).  
\label{eqApp:TabFu}
\eeq
To derive Eq. (\ref{eq:calcQ}), we have also used a condition 
\beq
\frac1{4\pi}\oint_S k^\gamma A_\gamma \Fabu dS_\albe
\,=\,0
\label{eq:epotconst}
\eeq
on the boundary sphere $S$ to determine the constant of 
the electric potential $\PhiE$ in Eq.~(\ref{eq:PhiE}).  
Eq.~(\ref{eq:calcQ}) implies that $Q(0)$ does not 
depend on the sphere $S$ as long as it encloses
all black holes and neutron stars, because in the region 
where the sphere $S$ is located, the integrand of the volume 
integrals in Eq.~(\ref{eq:calcQ}) vanishes 
when the Einstein and Maxwell equations are satisfied.

Next, the variation $\dl Q := dQ/d\lambda$ 
in the Noether charge is evaluated in terms of 
perturbations of the baryon mass, entropy, circulation and electric 
current of each fluid element, and the surface areas and charges of 
the black holes.  The detailed calculation is described in \cite{UGM10}:  the variation 
$\dl Q$ becomes 
\beqn
\delta Q  
&=& 
\int_\Sigma \left\{\,\frac{T}{u^t} \Dl (s \, \rho u^\alpha\, \dSa)
\right.
\,+\,\frac{h-Ts}{u^t}\Dl(\rho u^\alpha\, \dSa)
\,+\,v^\beta \Dl(hu_\beta\,\rho u^\alpha\,\dSa) 
\,-\, A_\beta k^\beta \, \Dl (j^\alpha \dSa)
\nonumber \\ 
&&\left.\phantom{\frac12}
\,-\,(j^\alpha k^\beta - j^\beta k^\alpha )
\Dl A_\beta \,\dSa
\,\right\}
\,+\,\sum_i \left(\frac1{8\pi}\kappa_i \dl {\cal A}_i
\,+\, \PhiEi \,\dl \QEi \right), 
\label{eq:1stlaw_org}
\eeqn
where $T$ is the temperature, $h$ is the relativistic specific enthalpy, 
and $v^\alpha$ is the spatial velocity defined by 
the following decomposition of the 4-velocity $u^\alpha$ 
with respect to the helical vector $k^\alpha$: 
\beq
u^\alpha = u^t(k^\alpha + v^\alpha) 
\quad\mbox{with}\quad 
v^\alpha\na_\alpha t = 0 
\quad\mbox{and}\quad 
u^t=u^\alpha \na_\alpha t.
\label{eq:4v_decomp}
\eeq 
The electric charge of each black hole $\QEi$ 
is defined by 
\beq
\QEi := \frac1{4\pi}\oint_{{\cal B}i} \Fabu dS_\albe, 
\label{eq:QEBH}
\eeq
which is related to the total electric charge of 
the system $\QE$ by Stokes' theorem:  
\beq
\QE
:= \frac1{4\pi}\oint_{S} \Fabu dS_\albe. 
\,=\,
\int_\Sigma j^\alpha \dSa
\,+\,\sum_i \QEi .
\label{eq:QEvol}
\eeq
Note that the $\PhiEi$, defined on each ${\cal B}_i$ by 
\beq
\PhiEi = -A^\alpha k_\alpha = \PhiE + C,
\eeq
is constant.  
In the case of stationary and axisymmetric spacetimes, 
the mass variation formula derived by Carter \cite{Carter73,Carter79} 
can be derived from Eq.~(\ref{eq:1stlaw_org}).

We can now verify that $Q(\lambda)$ is 
independent of the location of the 2-surface $S$, 
because the charge $Q(\lambda)$ at 
$\lambda=0$ is shown to be independent of $S$, and the variation 
formula (\ref{eq:1stlaw_org}) implies that $dQ/d\lambda = \dl Q$ 
is independent of $S$ as long as it encloses the fluid 
and black holes.  Moreover, 
in \cite{UGM10} we have shown that the difference, 
\beq
\dl \Big(Q \,-\, \sum_i Q_i 
\,-\,\frac1{4\pi}\int_{\pa \Sigma}
k^\gamma A_\gamma F^{\alpha\beta} dS_{\alpha\beta} \Big) 
\,=\,
\dl Q
\,-\, 
\sum_i\Big(
\frac1{8\pi}\kappa_i \dl{\cal A}_i 
\,+\,\PhiEi \,\dl \QEi \Big), 
\label{eq:dlQ_inv}
\eeq
is invariant under a gauge transformation that respects the symmetry, 
and hence we verify that so is $\dl Q$.

\subsection{Application of the first law to solution sequences in equilibrium}

When  inspiraling binary systems, or isolated neutron stars are evolving 
adiabatically -- in a timescale much longer than the dynamical timescale -- 
they may be modeled by a sequence of solutions in equilibrium.  
When the first law for a stationary and axisymmetric perfect fluid spacetime 
is applied to a sequence of rotating neutron star solutions whose local 
changes of rest mass and entropy are constant, the first law becomes 
$\dl M = \Omega \dl J$.  This is a condition for applying the turning point 
theorem \cite{turning1} to determine the stability of solutions.  
In this section, we consider an application of the first law 
(\ref{eq:1stlaw_org}) to a sequence of helically symmetric  solutions 
that model inspiraling binary black holes and/or neutron stars, 
assuming that the neutron star matter is a perfect conductor.

For a perfectly conducting medium, the ideal MHD condition, 
\beq
\Fabd u^\beta \,=\, 0,
\label{eq:idealMHD}
\eeq
is satisfied, and hence its curl is written 
\beq
\Lie_u \Fabd \,=\, 0, 
\label{eq:conservMflux}
\eeq
which is the magnetic flux conservation law, Alfven's law.  
For each equilibrium solution, rest mass 
and entropy are conserved: 
\beq
\Lie_u(\rho\sqrt{-g})=0, \quad
\Lie_u s = 0.
\label{eq:conservrhos}
\eeq
When we consider a sequence of solutions along which the rest mass, entropy, 
and magnetic flux are all conserved, the perturbed conservation laws 
corresponding to the above Eqs. (\ref{eq:conservMflux}) and (\ref{eq:conservrhos}) 
have first integrals: 
\beq
\Dl (\rho\sqrt{-g}) = 0, 
\quad
\Dl s = 0, 
\quad \mbox{and} \quad
\Dl \Fabd = 0.
\label{eq:pconsrhosF}
\eeq
For black holes, we may assume that areas and charges are constant. 
With these assumptions, the first law (\ref{eq:1stlaw_org}) is rewritten 
\beq
\delta Q  
\,=\,
\int_\Sigma \left\{\, v^\beta \Dl(hu_\beta\,\rho u^\alpha\,\dSa) 
\,-\, A_\beta k^\beta \, \Dl (j^\alpha \dSa)
\,\right\}.
\label{eq:1stlaw_1}
\eeq
Here $\Dl \Fabd =0$ with $\Fabd = (dA)_\albe $, 
$\Dl (dA)_\albe = (d \Dl A)_\albe $, whence the Poincar\'e lemma implies that there exists a function $\Psi$ such that 
$\Dl A_\alpha = \na_\alpha \Psi$.  From a substitution of this to 
the last term in the volume integral of Eq. (\ref{eq:1stlaw_org}), the latter
is shown to vanish.  
The remaining terms in Eq. (\ref{eq:1stlaw_1}) are related to the circulation 
of the magnetized flow.  

For a perfect fluid without magnetic fields, the conservation of circulation 
is written as $\Lie_u \hat \omega_\albe = 0$ where $\hat \omega_\albe := (d(hu))_\albe$ is the vorticity tensor. 
When  circulation is conserved along a sequence of solutions, 
it implies that $\Dl \hat \omega_\albe = 0$, and hence the first term in the integral 
of Eq. (\ref{eq:1stlaw_1}) vanishes \cite{Friedman:2001pf}.  
In general, there is no such conservation law for the circulation 
of magnetized flow.  However, Bekenstein and Oron \cite{Bekenstein:2000sf} 
(Tarapov and Gorskii \cite{TarapovGorskii} for Newtonian MHD) have developed 
a formulation of  ideal MHD, 
in which a generalized circulation of magnetized flow is conserved.  Their theory is based on a Lagrangian in which the interaction term $A_\alpha j^\alpha$ in 
Eq. (\ref{eq:Lag}) is replaced by the term $\Fabd \rho u^\alpha q^\beta$, where the 
vector $q^\alpha$ is a Lagrange multiplier enforcing the perfect conductivity condition $\Fabd u^\alpha =0$. Variation with respect to $A_\alpha$ yields the Maxwell equations with a current of the form 
\beq
j^\alpha \,=\, \na_\beta(\rho u^\alpha q^\beta - \rho u^\beta q^\alpha).  
\label{eq:BBcur}
\eeq

When a vector $q^\alpha$ is found to give the current (\ref{eq:BBcur}), 
the Lorenz force term in the MHD-Euler equation 
\beq
\label{eq:MHD-Euler}
u^\beta (d(hu))_\beal \,=\, \frac1{\rho} \Fabd j^\beta
\eeq
can be absorbed into the left hand side,  bringing eq. (\ref{eq:MHD-Euler}) 
in the canonical form 
\beq
u^\beta \omega_\beal \,=\, 0, 
\label{eq:MHD-Euler_BB}
\eeq
where
\beq
\omega_\albe := (dw)_\albe, 
\quad 
w_\alpha \,:=\, hu_\alpha \,+\,\eta_\alpha,
\quad
\eta_\alpha:=\Fabd q^\beta,  
\eeq
because of a relation, 
\beq
\frac1{\rho}\Fabd j^\beta 
\,=\,
\frac1{\rho}\Fabd \left[\,\Lie_q (\rho u^\beta) 
\,+\, \rho u^\beta \na_\gamma q^\gamma\,\right]
\,=\, (d\eta)_\albe u^\beta.  
\eeq
The one-from $w_\alpha$ and its exterior derivative $\omega_{\alpha \beta}$ can be  respectively regarded as a canonical momentum of a magneto-fluid element  and a  generalized vorticity. The MHD-Euler equation (\ref{eq:MHD-Euler_BB}) in this case  implies 
 conservation of  circulation for the magnetized flow, because 
its curl, with $d\omega = 0$, yields 
\beq 
\Lie_u \omega_\albe\,=\,0, 
\eeq
and the Lagrangian perturbation of this conservation law has a first integral 
\beq
\Dl \omega_\albe\,=\,0.
\label{eq:pconsMagcirc}
\eeq

Substituting the current (\ref{eq:BBcur}) to the first law (\ref{eq:1stlaw_org}), 
we have 
\beqn
\delta Q
&=& 
\int_\Sigma \left\{\,\frac{T}{u^t}  \Dl dS
\,+\,\frac{h-Ts}{u^t}\Dl dM_{\rm B}
\,+\,v^\alpha \Dl dC_\alpha
\,-\,v^\beta q^\gamma\Dl F_{\beta\gamma}\,dM_{\rm B}
\,-\,(j^\alpha k^\beta - j^\beta k^\alpha )
\Dl A_\beta \,\dSa
\,\right\}
\nonumber \\ 
&+&\sum_i \left(\frac1{8\pi}\kappa_i \dl {\cal A}_i
\,+\, \PhiEi \,\dl \QEi \right), 
\label{eq:1stlaw_BBcur2}
\eeqn
where we introduced the  notation
\beq
dM_{\rm B}:=\rho u^\alpha \dSa, 
\quad
dS:= s\,dM_{\rm B}, 
\quad
dC_\alpha:= (h u_\alpha +\eta_\alpha) dM_{\rm B}.
\eeq
Applying the first law (\ref{eq:1stlaw_BBcur2}) to a sequence 
of solutions along which the quantities are conserved as in
Eq. (\ref{eq:pconsrhosF}) and the circulation of magnetized 
flow is conserved as in Eq. (\ref{eq:pconsMagcirc}), the first law becomes 
\beq
\dl Q \,=\, 0, 
\eeq
or for asymptotically flat systems, such as in the post-Newtonian 
approximation, $\dl Q = \dl M - \Omega \dl J =0$.

\section{A formulation for computing  magnetized neutron star equilibria}

To compute solutions of  single or binary compact objects in equilibrium, 
we apply a finite difference scheme, or a pseudo-spectral method, 
to a system of basic equations and numerically solve the system.  
Those equations include the Einstein equation, the Maxwell equations, 
the MHD-Euler equation, and the baryon mass conservation equation.  
Here, we assume that the flow is homentropic, and hence the neutron-star matter is described by  
a one-parameter EOS.  

For the gravitational field, the 
Isenberg - Wilson - Mathews formulation, the 
waveless formulation, or a full set of Einstein's equations 
assuming helical symmetry may be used.  
These formulations are based on a 3+1 decomposition of  spacetime, 
and reliable numerical methods to solve these equations have been 
developed \cite{BNSCF,UryuCFWL,BHNS_QE}.  We expect that 
the Maxwell equations can be solved using analogous formulations 
and applying one of the above numerical methods.  

However, for the MHD-Euler equation, when  stationarity or 
helical symmetry is imposed, it is no longer an evolution equation, 
and as a result it is difficult to integrate numerically.  
In self-consistent field methods \cite{BNSCF,UryuCFWL,BHNS_QE,PriceMarkakisFriedman2009}, 
an equilibrium solution is computed using a first integral of the (MHD-) Euler equation which exists when the flow is assumed to be either corotational or irrotational.  
Therefore, finding the first integral of the MHD-Euler equation
is a key, and also a restriction, for computing equilibrium solutions 
considered in the previous section successfully.  

If we assume that the orbit of the stars is closed and quasi-circular, then the system appears stationary in a frame rotating with frequency equal to the orbital frequency $\Omega$, and there exists an approximate helical Killing vector $k^\alpha$, given by eq. \eqref{eq:hkv}, that Lie-derives all  variables in \eqref{eq:Qvars}. The Cartan identity implies
\beq
k^\beta(dw)_{\beta \alpha}=\Lie_{k} w_\alpha -
\na_\alpha(k^{\gamma}w_\gamma) 
\label{eq:cartan_k}
\eeq
It should be noticed that $\Lie_k w_\alpha \neq 0$ in general, since $q^\alpha$ is not 
an observable quantity. One has  
\beq 
\Lie_{k} w_\alpha 
\,=\, \Lie_{k} \eta_\alpha 
\,=\, \Fabd \Lie_k q^\beta.  
\label{eq:liekw}
\eeq
(In fact, if one assumes 
$\Lie_k w_\alpha = 0$ for a corotational flow, no Lorenz force 
is exerted on the matter). As shown below (see also \cite{GourgoulhonRelativisticFluids}), the Cartan identity  \eqref{eq:cartan_k} leads quickly to conserved quantities  when certain assumptions for the flow are made.

\subsection{Irrotational magneto-flow}
If one  assumes that the magneto-flow is 
irrotational, i.e. described by a potential $\Phi$: 
\beq
w_\alpha=hu_\alpha \,+\,\eta_\alpha = \na_\alpha \Phi,  
\eeq
then the vorticity $\omega=dw$ vanishes and the MHD-Euler equation (\ref{eq:MHD-Euler_BB})
is automatically satisfied while the left hand side of eq. \eqref{eq:cartan_k}
 vanishes  as well, so that
\beq
\Lie_{k} w_\alpha -
\na_\alpha(k^{\gamma}w_\gamma) =0
\label{eq:cartan_heli_zero}
\eeq
 The above equation has a first integral iff the term $\Lie_{k} w_\alpha$ is the
gradient of some scalar function. Because of eq. \eqref{eq:liekw}, this integrability condition is written as $\Lie_{k} q^\beta F_\beal = \na_\alpha f$,  
or, using the Cartan identity, 
\beq
\Lie_{[k,q]} \Fabd \,=\, 0,
\label{eq:intcond}
\eeq
where $[k,q]^\alpha := \Lie_k q^\alpha$. Then,  eq. \eqref{eq:cartan_k}
implies
that the quantity
\beq
k^{\gamma}w_\gamma+f \,=\, {\cal E}   
\label{eq:1stint_irrot1}
\eeq
is constant throughout the magneto-fluid. 

In this case, we have four variables 
for the matter, two thermodynamic variables and two variables 
for the velocity fields, $\{h,p,\Phi,u^t\}$.  The set of equations 
to solve for these four variables is supplemented by a one-parameter EOS, $p=p(h)$, 
and the normalization of the 4-velocity $u_\alpha u^\alpha=-1$.  
The velocity potential $\Phi$ is obtained from the rest mass conservation 
equation which is an elliptic equation for $\Phi$ with Neumann boundary 
conditions on the stellar surface.  An additional degree of freedom may be obtained from 
the first integral \eqref{eq:1stint_irrot1}, if it exists.


It is not trivial to find a vector $q^\alpha$ that satisfies 
both the integrability condition (\ref{eq:intcond}) and the ideal MHD 
condition (\ref{eq:idealMHD}); $q^\alpha$ is not freely specifiable.  
Instead of solving for a $q^\alpha$ that satisfies both conditions, let us assume that
\beq
q^\alpha = q^t k^\alpha
\eeq
with $\Lie_k q^t $ regarded a function of $A_\alpha k^\alpha$.
If we select $q^\alpha$ as above, at least instantaneously 
on an initial hypersurface $\Sigma$, then the first integral is written  
\beq
k^{\alpha}w_\alpha
\,+\, \int \Lie_{k} q^t \,d(A_\alpha k^\alpha)  \,=\, {\cal E}
\label{eq:1stint_irrot}
\eeq
It is likely that solutions with neutron stars and strong magnetic fields 
computed from Eq.~(\ref{eq:1stint_irrot}) may not be strictly in equilibrium.  
Nonetheless, such solutions with strong magnetic fields that 
satisfy a set of hydrostationary equations will serve as interesting 
initial data sets for  numerical relativity merger simulations and 
for further studies of such strongly magnetized compact binaries. 

\subsection{Corotational magneto-flows}
If the flow is corotational, as is the case for a tidally locked  binary or a  rigidly rotating non-axisymmetric magnetar, then  the velocity field can be written it terms of the helical Killing vector \eqref{eq:hkv} as
\beq
u^\alpha=u^tk^{\alpha}
\label{eq:corot_flow}
\eeq 
In this case, the MHD-Euler equation \eqref{eq:MHD-Euler_BB}
implies that the left-hand side of the Cartan identity \label{eq:cartan_heli} \eqref{eq:cartan_k} vanishes, so that eq. \eqref{eq:cartan_heli_zero} again holds. The integrability condition of the latter equation then is formally identical to eq. \eqref{eq:intcond} and the first integral of  eq. \eqref{eq:cartan_heli_zero} again has the form of eq. \eqref{eq:1stint_irrot1}:
\beq
k^{\gamma}w_\gamma+f \,=\, {\cal E}   
\label{eq:1stint_corot1}
\eeq
In the corotational case, the perfect MHD condition \eqref{eq:idealMHD}  also admits a first integral:
\beq
k^{\gamma}A_\gamma =0
\label{eq:corot_flow_kdotA}
\eeq
This follows  from the Cartan identity,
\beq
k^\beta(dA)_{\beta \alpha}=\Lie_{k} A_\alpha -
\na_\alpha(k^{\gamma}A_\gamma), 
\label{eq:cartan_heli_A}
\eeq
 along with the facts that the term $k^\beta(dA)_{\beta \alpha}$ vanishes by virtue of eqs. \eqref{eq:corot_flow}
and \eqref{eq:idealMHD}, and that the term $\Lie_{k} A_\alpha$ can be made to vanish by a gauge transformation in $A_\alpha$ (c.f.  Appendix C in \cite{GMU}). One can then incorporate the first integrals 
\eqref{eq:1stint_corot1} and \eqref{eq:corot_flow_kdotA} in a self-consistent iteration scheme analogous to that outlined  for the irrotational case above, coupled to the Einstein-Maxwell equations, to construct equilibrium models of rigidly rotating configurations. A scheme based on eq. \eqref{eq:1stint_corot1}, with $f=0$, has been implemented in \cite{HMSU} for constructing non-magnetized triaxial (Jacobi) spheroids in general relativity and can be extended to  model magnetars that are triaxially deformed by their magnetic field.

\section{Discussion}

The thermodynamic laws derived for helically symmetric Einstein-Maxwell 
spacetimes with magnetized matter can be applied to a 
non-axisymmetirc single rotating star, or an axisymmetric 
rotating star beyond the stationary, axisymmetric 
\emph{and circular} spacetime.  Our project involves computing quasi-equilibrium configurations of magnetized compact objects with numerical codes based on self-consistent field methods. Such methods can be used to model 
not only binary compact objects, but also magnetars, as well as proto neutron stars
which are likely to have strong magnetic fields.  

\ack
We thank Brandon Carter, Yoshiharu Eriguchi, John  Friedman and Ichiro Oda for enlightening 
discussions and suggestions.
This work was supported by 
JSPS Grant-in-Aid for Scientific Research(C) 20540275, 
MEXT Grant-in-Aid for Scientific Research
on Innovative Area 20105004, 
NSF Grant PHY100155
and ANR grant 06-2-134423 \emph{M\'ethodes 
math\'ematiques pour la relativit\'e g\'en\'erale}.  
CM thanks the Greek State Scholarships Foundation for support during 
the early stages of this work and the Paris Observatory and 
the University of Wisconsin-Milwaukee for travel support.
KU and EG acknowledge support from the JSPS Invitation Fellowship 
for Research in Japan (Short-term) and the invitation program 
of foreign researchers at the Paris Observatory.

\end{document}

%% file: kigou.tex
\newcommand{\hB}{{\mathfrak B}}

\newcommand{\beq}{\begin{equation}} 
\newcommand{\eeq}{\end{equation}} 
\newcommand{\beqn}{\begin{eqnarray}} 
\newcommand{\eeqn}{\end{eqnarray}} 
\newcommand{\pa}{\partial}
\newcommand{\na}{\nabla}

\newcommand{\gabu}{g^{\alpha\beta}}
\newcommand{\gabd}{g_{\alpha\beta}}

\newcommand{\albe}{{\alpha\beta}}
\newcommand{\beal}{{\beta\alpha}}

\newcommand{\TabFu}{T_{\rm F}^{\alpha\beta}}
\newcommand{\TabF}{T_{{\rm F}}{}\!^\alpha{}\!_\beta}

\newcommand{\Gab}{G^\alpha{}\!_\beta}

\newcommand{\Fabu}{F^{\alpha\beta}}
\newcommand{\Fabd}{F_{\alpha\beta}}

\newcommand{\Fcdu}{F^{\gamma\delta}}
\newcommand{\Fcdd}{F_{\gamma\delta}}

\newcommand{\PhiE}{{\Phi^{\rm E}}}
\newcommand{\PhiEi}{{\Phi^{\rm E}_i}}
\newcommand{\QE}{Q^{\rm E}}
\newcommand{\QEi}{Q^{\rm E}_i}

\newcommand{\Hori}{{\cal H}^{\pm}}
\newcommand{\Horiplus}{{\cal H}^{+}}

\newcommand{\zD}{{\raise1.0ex\hbox{${}^{\ \circ}$}}\!\!\!\!\!D}
\newcommand{\alone}{{\raise0.5ex\hbox{${}^{\ 1}$}}\!\!\!\!\alpha}

\newcommand{\dSab}{dS_{\alpha\beta}}
\newcommand{\dSa}{dS_{\alpha}}

\newcommand{\dl}{\delta}

\newcommand{\Dl}{\Delta}

\newcommand{\Lie}{\mbox{\pounds}}
\newcommand{\Lag}{{\cal L}}

\newcommand{\nalam}{\mathrel{\raise0.9ex\hbox{$^\lambda$}\mkern-14mu
\lower0.0ex\hbox{$\nabla$}}}

\newcommand{\zeroD}{{\raise1.0ex\hbox{${}^{\ \circ}$}}\!\!\!\!\!D}
\newcommand{\zLap}{{\raise1.0ex\hbox{${}^{\ \circ}$}}\!\!\!\!\Delta}
\newcommand{\zna}{\stackrel{{\circ}}{\nabla}}
\newcommand{\zS}{{\raise1.0ex\hbox{${}^{\ \circ}$}}\!\!\!\!\!S}

%% file: MBNS_proceedings.bbl
\section*{References}
\begin{thebibliography}{9}


\bibitem{2006csxs.book..547W} 
Woods P M, \& Thompson C 2006, 
In \textit{Compact stellar X-ray sources} ed W~Lewin and 
M~van der Klis, Cambridge Astrophysics Series, No.~39, 547, 
Cambridge University Press.


\bibitem{BNSmagnetized}
  Liu Y T, Shapiro S L, Etienne  Z B and Taniguchi K 2008,
  Phys.\ Rev.\  D {\bf 78}, 024012; 
%
\ Anderson M,  Hirschmann E W,  Lehner L,  Liebling S L, 
\ Motl P M,  Neilsen D,  Palenzuela C,  Tohline J E 2008,
  Phys.\ Rev.\ Lett.\  {\bf 100}, 191101. 
%
  Giacomazzo B, Rezzolla L and Baiotti L 2009,
  \textit{Preprint} gr-qc/0901.2722.


\bibitem{BD92D94}
  Blackburn J K and Detweiler S 1992, Phys.\ Rev.\ D {\bf 46}, 2318; 
~Detweiler S 1994, Phys.\ Rev.\ D {\bf 50}, 4929.

\bibitem{schild}
Schild A 1963, Phys.\ Rev.\ 131, 2762 

\bibitem{FU06GU07}
  Friedman J L and Uryu K 2006,
  Phys.\ Rev.\  D {\bf 73}, 104039;
  Glenz M M and Uryu K 2007,
  Phys.\ Rev.\  D {\bf 76}, 027501.

\bibitem{FRW} 
  Friedrich H, R\'acz I and Wald R M 1999, 
  Comm.\ Math.\ Phys.\ {\bf 204}, 691.



\bibitem{Carter73}
Carter B 1973,  
\textit{Black hole equilibrium states}, 
in \textit{Black Holes} \textit{(Les Astres Occlus)}, 
p.~57-214. 

\bibitem{Carter79}
Carter B 1979,  
\textit{The general theory of the mechanical, electromagnetic 
and thermodynamic properties of black holes}, 
in \textit{General Relativity: An Einstein centenary survey}, 
ed. S W Hawking \& W Israel,
p.~{294-369};
B~Carter, 
\textit{Generalised mass variation formula for a stationary 
axisymmetric star or black hole with surrounding accretion disc}, 
in \textit{Anneau d'accr\'etion sur les trous noires}, 
(Facult\'e des Sciences, Angers), 
ed. I~Moret-Bailly \& C~Latremoli\`ere,
p.~166-182;  
B.~Carter, 
\textit{Perfect fluid and magnetic field conservation laws 
in the theory of black hole accretion rings}, 
in \textit{Active galactic nuclei}.~(A79-50785 22-90) 
Cambridge, Cambridge University Press, 
ed. C Hazard \& S Mitton,
p.~273-300.

\bibitem{wald} Wald R M 1993, Phys.\ Rev.\ D. {\bf 48}, R3427 

\bibitem{iyerwald} Iyer V and  Wald R M 1995, Phys.\ Rev.\ D. {\bf 52}, 4430

\bibitem{UGM10}
Uryu K, Gourgoulhon E and Markakis C 2010, 
Phys. Rev. D \textbf{82}, 104054
\bibitem{turning1} Sorkin R 1981, Astrophys.\ J.\ {\bf 249}, 254.

\bibitem{Friedman:2001pf}
  Friedman J L, Uryu K and Shibata M (FUS) 2002, 
  Phys.\ Rev.\  D {\bf 65}, 064035 
  [Erratum-ibid.\  2004, D {\bf 70}, 129904]


%
\bibitem{Bekenstein:2000sf}
Bekenstein J D and Oron A 2000,
  Phys.\ Rev.\  E {\bf 62}, 5594;
%
 Bekenstein J D and Oron A 2001,
  Found.\ Phys.\  {\bf 31}, 895
%


\bibitem{TarapovGorskii}
Tarapov I E 1984, PMM U.S.S.R., 48, 275; 
~Gorskii V B 1986, PMM U.S.S.R., 50, 388.



\bibitem{BNSCF}
%
~Bonazzola S, Gourgoulhon E and Marck J-A 1999, 
Phys.\ Rev.\ Lett.\ 82, 892; Gourgoulhon E, Grandclement P, 
~Taniguchi K, Marck J-A, Bonazzola S 2001, Phys.\ Rev.\ D 63, 
064029; Taniguchi K and Gourgoulhon E 2002,
Phys.\ Rev.\ D 66, 104019; ibid. 2003, 68, 124025; 
%
%
  ~Bejger M, Gondek-Rosinska D, Gourgoulhon E, ~Haensel P, Taniguchi K and Zdunik J L 2005,
  Astron.\ Astrophys.\  {\bf 431}, 297;
%
%
  ~Taniguchi K and Shibata M 2010,
  Astrophys.\ J.\ Suppl.\  {\bf 188}, 187
%
\bibitem{UryuCFWL}
Uryu K and Eriguchi Y 2000, Phys.\ Rev.\ D \textbf{61}, 124023; 
~Uryu K, Shibata M and Eriguchi Y 2000, Phys.\ Rev.\ D \textbf{62},
104015;
  Uryu K, Limousin F, Friedman J L, Gourgoulhon E and Shibata M 2006,
  Phys.\ Rev.\  Lett.\ {\bf 97}, 171101; 
  Uryu K, Limousin F, Friedman J L, Gourgoulhon E and Shibata M 2009,
  Phys.\ Rev.\  D {\bf 80}, 124004.


\bibitem{BHNS_QE}
%
  Taniguchi K, Baumgarte T W, Faber J A and Shapiro S L 2005,
  Phys.\ Rev.\ D {\bf 72}, 044008; 
%
 Taniguchi K, Baumgarte T W, Faber J A and Shapiro S L 2006,
  Phys.\ Rev.\  D {\bf 74}, 041502(R); 
%
  Taniguchi K, Baumgarte T W, Faber J A and Shapiro S L 2007,
  Phys.\ Rev.\  D {\bf 75}, 084005 (2007); 
%
%
 Taniguchi K, Baumgarte T W, Faber J A and Shapiro S L 2008,
  Phys.\ Rev.\  D {\bf 77}, 044003; 
%
Kyutoku K, Shibata M and Taniguchi K 2009, 
Phys.\ Rev.\  D {\bf 79}, 124018; 
%
  Foucart F, Kidder L E, Pfeiffer H P and Teukolsky S A 2008,
  Phys.\ Rev.\  D {\bf 77}, 124051. 

\bibitem{PriceMarkakisFriedman2009}
Price R H, Markakis C and Friedman J L 2009,    J. Math. Phys. \textbf{50}, 073505 (\textit{Preprint}        astro-ph.SR/0903.3074) 

\bibitem{GourgoulhonRelativisticFluids}
Gourgoulhon E 2006, \textit{An Introduction to Relativistic Hydrodynamics}, in “\textit{Stellar fluid dynamics and numerical simulations}, ed M Rieutord \& B Dubrulle, EAS Publications Series \textbf{21} p. 43-79 (\textit{Preprint} gr-qc/0603009)

\bibitem{OM68}
Ostriker J P and Mark J~W-K 1968, Astrophys.\ J.\ \textbf{151}, 1075;  
~Hachisu I 1986, Astrophys.\ J.\ Suppl.\ \textbf{62}, 461; ibid. 61, 479 (1986);
Komatsu H, Eriguchi Y and Hachisu I 1989, 
Mon.\ Not.\ Roy.\ Astron.\ Soc.\ \textbf{237}, 355 




\bibitem{GMU}
Gourgoulhon E, Markakis C, Uryu K and Eriguchi Y 2011,   
\textit{Magnetohydrodynamics in stationary and axisymmetric spacetimes: a fully covariant
approach},
\textit{Preprint} gr-qc/1101.3497  


\bibitem{HMSU}
Huang X, Markakis C, Sugiyama N and Uryu K 2008, Phys. Rev. D \textbf{78} 124023 (\textit{Preprint} gr-qc/0809.0673) 


\end{thebibliography}
